\documentclass[12pt]{article}

\makeatletter

\def\@authoraddress{}
\def\@title{}
\def\title#1{\gdef\@title{{\par\vskip-10pt\Large\bf
\baselineskip20pt\centering\ignorespaces\uppercase{#1}\vskip6pt}}%
\setcounter{table}{0}      \setcounter{figure}{0}
\setcounter{equation}{0}   \setcounter{section}{0}
\setcounter{subsection}{0} \setcounter{subsubsection}{0}
\setcounter{paragraph}{0}
}

\def\authors#1{\expandafter\def\expandafter\@authoraddress\expandafter
{\@authoraddress %
{\dimen0=-\prevdepth \advance\dimen0 by1.5\baselineskip
\nointerlineskip \centering
\vrule height\dimen0 width0pt\relax\ignorespaces\large\sc#1\par
}%
}%
}

\def\addresses#1{\expandafter\def\expandafter\@authoraddress\expandafter
{\@authoraddress{\nointerlineskip\vskip1pc
                 \footnotesize\it\centering\ignorespaces#1\par}}}

\def\nextaddress{\\[2.3pt]}

\def\@maketitle{%
\@title
\ifdim\prevdepth=-1000pt \prevdepth0pt\fi
\@authoraddress
}

\def\maketitle{\par
\begingroup
\let\cite\@bylinecite
\global\@topnum\z@ %
\@maketitle
\endgroup
\def\@thanks{}\def\@authoraddress{}\def\@title{}
}

\def\abstract{\par
\bgroup
\ifdim\prevdepth=-1000pt \prevdepth0pt\fi
\hsize\columnwidth
\leftskip=2em \rightskip\leftskip
\dimen0=-\prevdepth \advance\dimen0 by2pc \nointerlineskip
\noindent\vskip1.5\baselineskip\nointerlineskip\noindent\footnotesize\relax}

\newif\if@firststuff

\def\endabstract{\par
\nointerlineskip \vskip0pt
\noindent \par
\egroup
\hrule depth0pt width0pt
\global\everypar{\global\@firststufffalse}\global\@firststufftrue
}

\renewcommand\section{\@startsection {section}{1}{\z@}%
                                   {-3.5ex \@plus -1ex \@minus -.2ex}%
                                   {2.3ex \@plus.2ex}%
                                   {\normalfont\large\bfseries}}
\renewcommand\subsection{\@startsection{subsection}{2}{\z@}%
                                     {-3.25ex\@plus -1ex \@minus -.2ex}%
                                     {1.5ex \@plus .2ex}%
                                     {\normalfont\large\bfseries}}

\def\1ad{\mbox{\normalsize $^1$}}
\def\2ad{\mbox{\normalsize $^2$}}
\def\3ad{\mbox{\normalsize $^3$}}
\def\4ad{\mbox{\normalsize $^4$}}
\def\5ad{\mbox{\normalsize $^5$}}
\def\6ad{\mbox{\normalsize $^6$}}
\def\7ad{\mbox{\normalsize $^7$}}
\def\8ad{\mbox{\normalsize $^8$}}
\def\adref#1{\mbox{\normalsize $^{#1}$}}
\makeatother

\begin{document}
\raggedbottom

\title{Holography, Time and Quantum Mechanics}

\authors{Vijay
Balasubramanian,\adref{1}
  Jan de Boer,\adref{2}
  and Djordje Minic\adref{3}}

\addresses{\1ad David Rittenhouse Laboratories, University of
Pennsylvania, Philadelphia, PA 19104, U.S.A.
  \nextaddress \2ad Instituut voor Theoretische Fysica, Valckenierstraat 65, 1018XE Amsterdam, The Netherlands
  \nextaddress \3ad Institute for Particle Physics and Astrophysics, Department of Physics, Virginia Tech, Blacksburg, VA 24061, U.S.A.}

\maketitle

\begin{abstract}

In this talk we entertain the possibility that the synthesis of
general covariance and quantum mechanics requires an extension of
the basic kinematical setup of quantum mechanics. According to the
holographic principle, regions of spacetime bounded by a finite
area carry finite entropy. When we in addition assume that the
origin of the entropy is a finite dimensional Hilbert space, and
apply this to cosmological solutions using a suitable notion of
complementarity, we find as a consequence that gravitational
effects can lead to dynamical variation in the dimensionality of
such Hilbert spaces. This happens generally in cosmological
settings like our own universe.

\end{abstract}

\section{Introduction}

Quantum mechanics describes the intrinsic fluctuations of systems
with a finite number of particles.  While the state of a classical
system is described as a point in position-momentum phase space,
the wavefunction of a quantum mechanical system is a ray in
Hilbert space.     The synthesis of quantum mechanics with the
special theory of relativity forces a generalization -- the theory
must accommodate transitions between states with different
particle number~\cite{wein}.   In particular, Lorentz invariance
implies the existence of anti-particles.   This leads to field
theory, in which states are described as vectors in Fock space, a
direct sum of Hilbert spaces describing different numbers of
particles.

In this talk we discuss the possibility that the synthesis of
quantum mechanics and general covariance requires  further
innovations in the basic kinematical setup of quantum mechanics.
According to the holographic principle, regions of spacetime
bounded by a finite area are described by a system with {\it
finite} entropy \cite{holog}.   It has been suggested by many authors that this entropy is associated with an accessible Hilbert space of finite dimensionality.    We point out  that gravitational effects
can lead to dynamical variation in the dimensionality of such
Hilbert spaces. This would happen generally in time-dependent cosmological settings
including our own universe.

\section{Local Gravitational Entropy and Hilbert Space Evolution}

It has been known for over 25 years that black holes have an
entropy proportional to surface area suggesting a deep connection
between the dynamics of gravity and thermodynamics~\cite{bhent}.
It has been proposed that the scaling of entropy with surface area
rather than volume is a general feature of quantum gravity dubbed
the {\it holographic principle}~\cite{holog}, wherein the dimension of the Hilbert space
required to describe a system enclosed by a
closed surface with area $A$ is bounded by $\exp{S}$ with $S =
A/4G_N$.   This suggests a breakdown of spacetime locality since
the proposed entropy is not extensive in the volume enclosed.  A
refinement of the proposal says that $\exp S$ is the entropy
encoded on converging lightsheets emerging from the closed
bounding surface~\cite{bousso}.

What does this proposal mean operationally?   Perhaps the most
remarkable aspect of it is that the entropy is finite, suggesting a
{\it finite} dimensional Hilbert space of excitations describing
the interior of a region bounded by a surface of area
$A$.\footnote{Note that finite entropy does not always imply a
finite dimensional Hilbert space.   In quantum mechanics the entropy
associated to a density matrix $\rho$ is $S=-{\rm tr}(\rho \log
\rho)$, and one can have finite $S$ and an infinite dimensional
Hilbert space, as is well-known from quantum statistical
mechanics.   Nevertheless, a finite $S$ typically implies that a finite number of states is operationally accessible.  In closed systems that are not coupled to an external heat bath (e.g., the universe as a whole) this implies a finite dimensional Hilbert space.  We will assume that finite gravitational entropy generally implies a finite dimensional accessible Hilbert space, but this
assumption clearly requires careful scrutiny.}   The cosmological
implications are particularly intriguing since the finite entropy
of, say, the horizon seen by inertial observers in de Sitter space
suggests that, operationally, any observer requires a finite
dimensional Hilbert space to describe his observable
universe~\cite{gibbons,banksetc,ds,strom,bdbm1,bdbm2}.\footnote{This finite dimensional Hilbert space is sufficient to describe the entire universe if there is also a notion of
complementarity, which states that the degrees of freedom behind
the horizon are the same as the ones seen by the observer
\cite{complementarity}. Whether there is indeed such a
notion is still a matter of debate, see e.g.
\cite{maldacenastrings}}   In contrast, in
conventional quantum mechanics, a single harmonic oscillator has
an infinite dimensional Hilbert space.

What are the implications of a local notion of holography
associated to local observers? Useful insights are gained by
invoking the equivalence principle and studying accelerated
observers.  Indeed, Jacobson has shown that assuming that the
thermodynamic laws of horizon mechanics apply to Rindler horizons
implies Einstein's equations~\cite{jacob}. Specifically, consider
families of accelerated local observers and assume that their
Rindler horizons enjoy a first law of the form
\begin{equation}
\delta E = T \delta S \, ,
\end{equation}
where  $\delta E = \int T_{\mu \nu} k^{\mu} k^{\nu} d \lambda$ is
the energy flux of matter through the horizon, $T$ is the local
Unruh temperature, $\delta S$ is a change in entropy, and
$\lambda$ is the affine parameter of the horizon.   Assume a
second law relating entropy and area
\begin{equation}
\delta S = \delta A  / 4 G_N\, .
\end{equation}
Here $\delta A = \int \theta d\lambda$ is the change in area of
the Rindler horizon, $\theta$ being the expansion of the horizon
generators.  Combining these equations with the Raychaudhuri
equation (here we drop the $\theta^2$ term as well as the term
involving the shear and $k^\mu$ is the tangent vector to the
horizon generators)
\begin{equation}
{d \theta \over d \lambda} = R_{\mu \nu} k^{\mu} k^{\nu} \, , \label{ray}
\end{equation}
yields  Einstein's gravitational field equations, remarkably
relating thermodynamic equations of state to dynamical equations
of motion~\cite{jacob}.   Let us now assume that the entropy is
determined by the dimensionality  of the accessible Hilbert space
$dim {\cal{H}} = \exp(S)$.  Then the Raychaudhuri equation relates
spacetime curvature to a  change in the  dimensionality of the
Hilbert space.

Given our assumptions, such changes in the dimensionality of
Hilbert spaces are  generic consequences of the synthesis of
general covariance and the usual interpretation of entropy in
quantum mechanics. Consider, for example, any asymptotically ${\rm
AdS}_5$ space ${\cal M}$ and the duality between string theory on
${\cal M} \times S^5$ and the corresponding deformation of the
superconformal $SU(N)$ Yang-Mills theory on $S^3 \times
R$~\cite{adscft}.  ${\cal M}$, being infinite in size, requires an
infinite dimensional Hilbert space to describe it.   Suppose we
cut off ${\cal M}$ at a finite radius -- the finite bounding area
then implies that a finite dimensional Hilbert space is needed to
describe the interior volume.   Indeed, the dual description of
the interior region is obtained by cutting off the CFT at a
certain energy scale, thereby obtaining a finite dimensional
Hilbert space~\cite{susswitt}.   In fact, the renormalization
group (RG) equation for the field theory dual to ${\cal M} \times
S^5$ is identified with the equations of motion governing radial
flow of the spacetime solution with the affine parameter $\lambda$
along the flow identified with the RG scale
$\Lambda$~\cite{holorg}.    The Raychaudhuri equation 
(of which(\ref{ray}) is a special case) 
applied to the radial flow of
surfaces that bound regions of ${\cal M}$ translates into an
equation for the RG evolution of the trace of the stress tensor of
the dual field theory~\cite{holorg}.    This quantity can be
interpreted as measuring the number of accessible degrees of
freedom $c$.  Coupled with weak energy condition, the Raychaudhuri
equation expresses the monotonicity of the flow of $c$: $c_{UV} >
c_{IR}$~\cite{holorg}.  In this example, radial flows in  ${\cal
M}$ are directly related via the renormalization group to changing
dimensionality of the accessible Hilbert space.

In Jacobson's discussion of Rindler horizons the change in the
dimensionality of the accessible Hilbert space occurs because
matter flows through the horizon.   While it is surprising that
the change in entropy is reflected in horizon area, it is natural
that more states are necessary to describe the space behind the
horizon after additional matter falls in.   We might say that the
dimensionality of the Hilbert space whose entropy is encoded in
horizon area changes because there is a flow of degrees of freedom
from one side of the horizon to the other.    In the AdS example
the dimension of the accessible Hilbert space changes with radial
flow because we explicitly choose to examine smaller regions of
space, requiring fewer degrees of freedom.  By contrast, below we
will demonstrate that in a cosmological setting the intrinsic
dynamics of gravity, rather than a choice of surface, can change
the dimension of the Hilbert space accessible to an inertial
observer.

\section{Holography, Cosmology and Entropy}

In the presence of a positive cosmological constant $\Lambda$, the
maximally symmetric solution to Einstein's equations is de Sitter
space.   Because of the rapid expansion of this universe, inertial
observers are surrounded by a cosmological horizon which hides
phenomena occurring in the region beyond.    This cosmological
horizon has a finite and constant area  leading to an entropy $S =
A/4G \sim \Lambda^{-(n-1)/2}/G_N$ for (n+1)-dimensional de Sitter
space~\cite{gibbons}. Many interpretations of de Sitter entropy
have been suggested including: (a) it measures the size of the
Hilbert space of the entire universe via ${\rm dim}\,H=e^S$, (b)
it measures the dimension of the Hilbert space of states that are
hidden behind the horizon, (c) it measures the size of the Hilbert
space of excitations that the inertial observer can interact with.
(See~\cite{banksetc} and references
in~\cite{ds,strom,bdbm1,bdbm2}.)   In any of these interpretations
the relevant Hilbert space is finite dimensional.\footnote{Another interpretation was proposed in  the context of a holographic dual description of de Sitter space in terms of an entangled state in a product of two (unconventional) conformal field
theories associated with the two temporal infinities of de Sitter
space~\cite{bdbm2}.   Here the entropy of the cosmological horizon of one de Sitter observer arose by tracing over the part of the entangled state that is inaccessible to the observer.}

An interesting perspective on the number of degrees of freedom required to describe asymptotically de Sitter spaces also arises if there is a holographic duality between such spaces and (deformations of) a Euclidean conformal field theory~\cite{ds}.  If there is such a duality, one would expect a relationship between  RG flow of the field theory dual and {\it time}
evolution in spacetime~\cite{strom,bdbm1}.  As always, in a theory of gravity,
the bulk Hamiltonian is zero, and so this is really a map between
a holographic RG equation and the Hamiltonian constraint of the
bulk gravitation theory, which at the quantum mechanical level
becomes the Wheeler-de Witt equation~\cite{wdw}.

We can put some flesh on this proposal by following the treatment of holographic RG flows in
asymptotically AdS spaces~\cite{holorg}. We fix the gauge so that the bulk metric
can be written as
\begin{equation}
ds^2 = -dt^2 + g_{ij}dx^i dx^j.
\end{equation}
The Hamiltonian constraint reads
\begin{equation}
{\cal{H}} =0,
\end{equation}
where in the case of 4d bulk gravity
\begin{equation}
{\cal{H}}= (\pi^{ij} \pi_{ij} - \frac{1}{2} \pi^{i}_{i} \pi^{j}_{j})
+\frac{1}{2} \pi_{I} G^{IJ}\pi_{J} + {\cal{L}}.
\end{equation}
Here $\pi_{ij}$ and $\pi_{I}$ are the canonical momenta conjugate to
$g^{ij}$ and $\phi^I$ ($\phi^I$ denotes some background test scalar
fields). ${\cal{L}}$ is a local Lagrangian density and
$G^{IJ}$ denotes the metric on the space of background scalar fields.

As in the context of the AdS/CFT duality~\cite{holorg}, the Hamiltonian
constraint can be formally rewritten as a renormalization group equation for
the dual RG flow
\begin{equation}
\frac{1}{\sqrt{g}}( \frac{1}{2} (g^{ij}
{\frac{\delta S}{\delta g^{ij}}})^2
-{\frac{\delta S}{\delta g^{ij}}}{\frac{\delta S}{\delta g_{ij}}}
-\frac{1}{2} G^{IJ}
\frac{\delta S}{\delta \phi^{I}} \frac{\delta S}{\delta \phi^{I}})
= \sqrt{g} {\cal{L}},
\end{equation}
provided the local 4d action $S$ can be separated into a
local and a non-local piece
\begin{equation}
S(g, \phi) = S_{loc}(g, \phi) + \Gamma (g, \phi).
\end{equation}
Indeed, in that case the Hamiltonian constraint can be formally rewritten as a
Callan-Symanzik renormalization group equation
\begin{equation}
\frac{1}{\sqrt{g}}( g^{ij}
{\frac{\delta }{\delta g^{ij}}} - \beta^I \frac{\delta}{\delta
\phi^I}) \Gamma = HO,
\end{equation}
where $HO$ denotes higher derivative terms.
Here the ``beta-function" is defined (in analogy with the
AdS situation) to be $\beta^I = \partial_{A}
\phi^{I}$,
where $A$ denotes the cut-off of the putative dual Euclidean theory.

Note that the separation of the local and non-local pieces in the 4d action
is completely analogous to the way the many-fingered physical time is
extracted from the Wheeler-de-Witt equation in the context of quantum
cosmology~\cite{wdw,banks1}.
There the non-WKB part of the wave function satisfies the
Tomonaga-Schwinger equation, which reduces to the usual time dependent
Schrodinger equation for global time variations~\cite{banks1}.

Some field theories have a "c-function" which measures the number of accessible degrees of freedom and which decreases during RG flow. Following the AdS literature~\cite{holorg},  we proposed a holographic c-function for asymptotically de Sitter spaces that is related to the trace of the Brown-York stress tensor~\cite{by} evaluated on equal time surfaces~\cite{bdbm1}. When the spacetime is four dimensional we would have
\begin{equation}
c \sim \frac{1}{G \theta^2},
\end{equation}
where $\theta$ is the trace of the extrinsic curvature of equal time surfaces.\footnote{An alternative c-function, based on the fact that generic perturbations to de Sitter "lengthen" the Penrose diagram has been given in \cite{myers}.}   Note that the trace of the Brown-York stress tensor turns out to be
\begin{equation} 
<T^{i}_{i}> \sim \theta,
\end{equation}
up to some terms constructed from local intrinsic curvature invariants
of equal time surfaces.    Hence the  the RG equation of a putative CFT dual to an asymptotically dS space would be given by
\begin{equation}
<T^{i}_{i}> \sim \frac{d \Gamma}{d A} = \beta^I \frac{\partial \Gamma}
{\partial \phi^I}.
\end{equation}
The Raychauduri equation then implies the monotonicity of the trace of the
Brown-York stress tensor
\begin{equation}
    \frac{d \theta}{dt} \le 0,
\end{equation}
as long as a form of the weak positive energy condition is satisfied
by the background test scalar fields.  This in turn guarantees monotonic time evolution
of the proposed  holographic $c$-function, suggesting that the number of degrees of freedom required to describe an asymptotically de Sitter space can change monotonically with time.

To study this we can also directly examine the area of the cosmological horizon in the generic situation of  a scalar field coupled to gravity in $n+1$
dimensions. Below we will show that this system admits solutions
which are expanding/contracting universes in which inertial
observers see a horizon whose size changes with time.  Therefore
the associated quantum mechanical Hilbert spaces must be changing
in dimension. The equations of motion derived from the action
\begin{equation}
S = {1 \over 16 \pi G_N} \int d^{n+1}x \sqrt{-g} \left[R - n(n-1) g^{ij} \partial_i\phi \partial_j\phi - n(n-1) V(\phi) \right],
\end{equation}
with a metric ansatz
\begin{equation}
ds^2 = -dt^2 + a(t)^2 \, d\Omega^2
\end{equation}
(where $d\Omega^2$ is the metric on the unit n-sphere) imply the Friedmann equations
\begin{equation}
{1 + \dot{a}^2 \over a^2} = \dot{\phi}^2 + V(\phi);
{\ddot{a} \over a} = -(n-1) \dot{\phi}^2 + V(\phi).
\end{equation}
The vacuum energy of the scalar field acts as a cosmological
constant $\Lambda$ and we a seek a potential and solutions that
interpolate between two values of $\Lambda$, implying that $a(t)
\rightarrow e^{-bt}$ and $a(t) \rightarrow e^{c t}$ as $t
\rightarrow \mp \infty$ respectively.   Positivity of
$\dot{\phi}^2$ implies
\begin{equation}
1 + \dot{a}^2 - a \ddot{a} \geq 0 \, .
\label{pos}
\end{equation}
It is easy to find a scale factor $a(t)$ and associated potentials
that meet these criteria.   In order to determine the size of the
horizon seen by an inertial observer we introduce conformal time
$\eta(t) = \int_{-\infty}^t dt/a(t)$ in terms of which
\begin{equation}
ds^2 = a(\eta)^2 \left[-d\eta^2 + d\theta^2 + \sin^2\theta \, d\Omega_{n-1}^2 \right] \, .
\end{equation}
The cosmological horizon is obtained by following null geodesics
emanating from $\eta = \theta = 0$.   This leads to a horizon area
and entropy that depends on conformal time as
\begin{equation}
S(\eta) = {[a(\eta) \sin\eta]^{n-1} {\rm vol}(\hat{S}_{n-1}) \over 4 G_N},
\label{entropy}
\end{equation}
where $\hat{S}_{n-1}$ is the unit (n-1)-sphere.   Interestingly
this entropy evolves monotonically just like a field theory
c-function at least  up to conformal times of $\eta = \pi/2$.
Indeed (\ref{pos}) implies $\eta(t) \geq \pi/2 + \arctan\dot{a}$,
which in turns implies that $\partial S /\partial \eta \geq 0$.
The expression for the entropy (\ref{entropy}) may be contrasted
with the proposed holographic c-function in de Sitter
space~\cite{myers}
\begin{equation}
c \sim \left( {a^2 \over 1 + \dot{a}^2} \right)^{(n-1)/2},
\end{equation}
in the context of a possible duality between de Sitter space and a
Euclidean conformal field theory~\cite{ds}.  The difference
between these expressions presumably lies in the fact that while
entropy measures the dimension of a Hilbert space, c counts
available degrees of freedom.

This example shows that in a cosmological context an inertial
observer may face a situation in which the accessible Hilbert
space changes in dimension with the passage of time.   Since the
universe as a whole is an isolated system, this change must be
attributed purely to the dynamical expansion of spacetime.   Given
that our universe probably passed through an era of inflation and
appears to be entering a new de Sitter phase~\cite{supernova}
these observations may have phenomenological relevance.

It is of course strange to have a Hilbert space whose dimension
changes with time. It cannot depend continuously on time, so if
one takes this seriously one is almost immediately led to a
discretization of space-time as well. A more conservative approach
might be to interpret the changing entropy as a change in the
entropy of a density matrix. Density matrices can easily depend on
continuous parameters, and in this way we don't have to deal with
the `quantized' dimension of a Hilbert space.

\section{Conclusion}

We have argued that the synthesis of quantum mechanics and general
covariance might require  innovations in the classic structures of
quantum mechanics.   According to the holographic principle,
regions of space of bounded surface area are described by finite
dimensional Hilbert spaces.   If so, we have pointed out that inertial
observers in a cosmological setting need to describe the world in
terms of Hilbert spaces whose dimensions vary in time and that our
universe may realize such a situation.     The 
mathematical framework for realizing such structures remains to be
uncovered, even though it is tempting to speculate that the
relevant mathematics might already have been discussed in the existing
literature~\cite{hilbert}.

\vspace{0.25in}
{\leftline {\bf Acknowledgments}}

We have learned about de Sitter space from many colleagues
including T.~Banks, P.~Berglund, R.~Bousso, R.~Dijkgraaf, W.~Fischler, P.~Ho\v{r}ava, T.~Hubsch, C.~Johnson, J.~Maldacena, D.~Marolf,  E.~Martinec, R.~Myers, M.~Li, E.~Mottola,  A.~Strominger, E.~Verlinde, H.~Verlinde, and
E.~Witten.  {\small V.B.} is supported by the DOE grant
DE-FG02-95ER40893.  {\small D. M.} is grateful to the organizers
of the 3rd Sakharov International Physics Conference for
hospitality and opportunity to present this work.

\end{document}